%% file: ms.tex
\documentclass{article}
\usepackage{bib/spconf}
\title{Domain Adaptation for Unknown Image Distortions in Instance Segmentation}
\name{Maximiliane Gruber, Fabian Brand, Alina Mosebach, J\"{u}rgen Seiler, and Andr\'{e} Kaup\thanks{We gratefully acknowledge support by the Bavarian Ministry for Economic Affairs, Regional Development and Energy (StMWi) under Grant No. DIK0179/02}} 
\address{Multimedia Communications and Signal Processing\\Friedrich-Alexander-Universit\"{a}t Erlangen-N\"{u}rnberg (FAU)}
\input{./settings/definitions.tex}

\newcommand\copyrighttext{%
	\footnotesize \textcopyright 2022 IEEE. Personal use of this material is permitted.
	Permission from IEEE must be obtained for all other uses, in any current or future 
	media, including reprinting/republishing this material for advertising or promotional 
	purposes, creating new collective works, for resale or redistribution to servers or 
	lists, or reuse of any copyrighted component of this work in other works. 
	DOI: \href{https://doi.org/10.1109/ICIP46576.2022.9897339}{10.1109/ICIP46576.2022.9897339}
}
\newcommand\copyrightnoticeOwn{%
	\begin{tikzpicture}[remember picture,overlay]
		\node[anchor=north,yshift=-10pt] at (current page.north) {\fbox{\parbox{\dimexpr\textwidth-\fboxsep-\fboxrule\relax}{\copyrighttext}}};
	\end{tikzpicture}%
	\vspace{-8mm}
}

\begin{document}
\hypersetup{pdfauthor={M. Gruber},pdftitle={2022-icip-gruber}}
\ninept
\maketitle
\copyrightnoticeOwn
%
\begin{acronym}
	\acro{A2D2}{Audi Autonomous Driving Dataset}
	\acro{AWGN}{additive white Gaussian noise}
	\acro{AP}{Average Precision}
	\acro{BDD100K}{Berkeley Deep Drive}
	\acro{CL}{Compression Level}
	\acro{CNN}{Convolutional Neural Network}
	\acro{DL}{Deep Learning}
	\acro{DNN}{Deep Neural Network}
	\acro{GAN}{Generative Adversarial Network}
	\acro{HEIF}{High Efficiency Image File Format}
	\acro{JPEG}{Joint Photographic Experts Group}
	\acro{R-CNN}{Region-based Convolutional Neural Network}
	\acro{mAP}{mean Average Precision}
	\acro{ML}{Machine Learning}
	\acro{MUNIT}{Multimodal UNpaired Image-to-image Translation}
	\acro{PSNR}{Peak Signal-to-Noise Ratio}
	\acro{QP}{Quantization Parameter}
	\acro{UNIT}{UNpaired Image-to-image Translation}
	\acro{OCR}{Optical Character Recognition}
	\acro{LPR}{License Plate Recognition}
\end{acronym}
%
\input{./texfiles/0_abstract.tex}
%
\input{./texfiles/1_introduction.tex}
%
\input{./texfiles/3_methods.tex}
%
\input{./texfiles/4_experiments.tex}

\input{./texfiles/5_conclusion.tex}
%
%
%
\bibliographystyle{./bib/IEEEbib}
\bibliography{ms.bib}
\end{document}

%% file: settings/definitions.tex
\usepackage[utf8]{inputenc}
\usepackage[T1]{fontenc}
\usepackage{amsmath, amsfonts, amssymb}
\usepackage[table]{xcolor} 
\usepackage{soul}
\usepackage{enumitem}
\usepackage{multirow}
\usepackage{lscape}
\usepackage{array}
\usepackage{booktabs} 
\usepackage[mode=buildnew]{standalone} 
\usepackage{graphicx}
\usepackage[caption=false]{subfig}
\usepackage{tikz,pgfplots,pgfplotstable}
\graphicspath{{images/}}
\usepackage[nolist]{acronym}
\usepackage{cite} 
\usepackage{siunitx}
\usepackage{hyperref}

\input{settings/macros}
\input{settings/tikz-settings}

%% file: settings/macros.tex
\providecommand{\myvec}[1]{\ensuremath{\boldsymbol{#1}}}       
\providecommand{\undistDom}{\ensuremath{X}}
\providecommand{\distDom}{\ensuremath{Y}}
\providecommand{\distFct}{\ensuremath{\mathcal{C}}}
\providecommand{\distFctLearned}{\ensuremath{\tilde{\distFct}}}
\providecommand{\generatorForward}{\ensuremath{G}}
\providecommand{\generatorBackward}{\ensuremath{F}}
\providecommand{\discriminatorDist}{\ensuremath{D_\distDom}}
\providecommand{\discriminatorUndist}{\ensuremath{D_\undistDom}}
\providecommand{\blurQP}{\ensuremath{\sigma_\mathrm{blur}}}
\providecommand{\noiseQP}{\ensuremath{\sigma_\mathrm{AWGN}}}
\providecommand{\jkQP}{\ensuremath{\mathrm{PSNR}}}
\providecommand{\jpegQP}{\ensuremath{\mathrm{CL}}}
\providecommand{\heifQP}{\ensuremath{\mathrm{QP}}}
\newcommand{\imageMagick}{\texttt{ImageMagick}}
\newcommand{\libheif}{\texttt{libheif}}
\newcommand{\detectron}{\texttt{Detectron2}}
\newcommand{\cyclegan}{\texttt{CycleGAN}}
\newcommand{\pixpix}{\texttt{pix2pix}}

%% file: settings/tikz-settings.tex
\usetikzlibrary{positioning,shadows,shapes}
\pgfplotsset{compat=1.17}
\usepgfplotslibrary{groupplots}
\pgfplotsset{every axis label/.append style={font=\small}}
\pgfplotsset{every tick label/.append style={font=\footnotesize}}
\pgfplotsset{every axis legend/.append style={font=\small}}
\pgfplotsset{every axis plot/.append style={line width=1.2pt}}
\pgfkeys{/pgf/number format/set thousands separator={\,}}
\pgfsetplotmarksize{1.5pt}
\tikzset{
 every picture/.style={>=latex},
 every node/.append style={font=\footnotesize},
 every pin/.append style={font=\tiny},
 every pin edge/.append style={shorten >=1pt, shorten <=.5pt},
}
\tikzstyle{syslinear} = [
 drop shadow={shadow xshift=.6mm,
              shadow yshift=-.6mm},
 fill=white,
 anchor=west,
 rectangle,
 draw=black,
 minimum height=5mm,
 minimum width=5mm,
 inner xsep=0.5em
]
\tikzstyle{sysnonlinear} = [
 drop shadow={shadow xshift=.8mm,
              shadow yshift=-.8mm},
 fill=white,
 double,
 anchor=west,
 rectangle,
 draw=black,
 minimum height=5mm,
 minimum width=5mm,
 inner xsep=0.5em
]
\tikzstyle{syssource} = [
 anchor=west,
 ellipse,
 draw=black,
 minimum height=1.5ex,
 minimum width=1.5em,
 inner xsep=0.5em
]
\tikzstyle{syssink} = [
 anchor=west,
 ellipse,
 draw=black,
 minimum height=1.5ex,
 minimum width=1.5em,
 inner xsep=0.5em
]
\tikzstyle{syssplit} = [
 circle,
 fill=black,
 draw=black,
 inner sep=1pt,
]
\tikzstyle{sysadd} = [
 draw,circle,inner sep=-1pt,
]
\tikzstyle{sysaddmod} = [
 draw,inner sep=-1pt,rectangle,inner xsep=-.3pt,inner ysep=-.2pt
]
\tikzstyle{sysmul} = [
 draw,circle,inner sep=0pt,
]
\tikzstyle{syscon} = [%
 draw,
 shape=circle,
 inner sep=0pt,
 minimum width=4pt,
 fill=white,
] 
\tikzstyle{system}   = [draw, shape=rectangle, inner sep=3pt, align=center] 
\tikzstyle{source}   = [draw, shape=ellipse, inner sep=3pt] 
\definecolor{myred}{HTML}{ff0000}
\definecolor{myblue}{HTML}{0000ff}
\definecolor{mygreen}{HTML}{4daf4a}
\definecolor{mylila}{HTML}{984ea3}
\definecolor{myorange}{HTML}{ff7f00}
\definecolor{myyellow}{HTML}{ffff33}
\definecolor{mybrown}{HTML}{a65628}
\definecolor{mypink}{HTML}{f781bf}
\definecolor{mygray}{HTML}{999999}
\definecolor{mylightgray}{HTML}{eeeeee}
\definecolor{mygridcolor}{HTML}{d9d9d9}

%% file: texfiles/0_abstract.tex
\begin{abstract}
	Data-driven techniques for machine vision heavily depend on the training data to sufficiently resemble the data occurring during test and application.
	However, in practice unknown distortion can lead to a domain gap between training and test data, 
	impeding the performance of a machine vision system.
	With our proposed approach this domain gap can be closed by unpaired learning of the pristine-to-distortion mapping function of the unknown distortion. 
	This learned mapping function may then be used to emulate the unknown distortion in the training data.
	Employing a fixed setup, our approach is independent from prior knowledge of the distortion.
	Within this work, we show that we can effectively learn unknown distortions at arbitrary strengths. 
	When applying our approach to instance segmentation in an autonomous driving scenario, we achieve results comparable to an oracle with knowledge of the distortion.
	An average gain in \acf{mAP} of up to \num{0.19} can be achieved.
\end{abstract}
\begin{keywords}
	Image Distortions, 
	Unpaired Image-to-Image Translation, 
	Unsupervised Domain Adaptation, 
	Instance Segmentation, 
	Autonomous Driving
\end{keywords}

%% file: texfiles/1_introduction.tex
\section{Introduction}
\label{sec:introduction}
Depending on the employed image acquisition setup and environmental conditions, images and videos contain different image distortions like blur, noise, low contrast, low resolution, coding artifacts, or combinations thereof.
In previous work, it has been shown that the performance of \ac{DNN}-based techniques for machine vision decreases if the input images or videos are subject to such distortions.
This negative impact has been shown for image classification~\cite{Dodge2016, Ghosh2018, Hendrycks2019, Pei2021, Endo2021}, semantic segmentation~\cite{Kamann2021}, object detection~\cite{Geirhos2018, Fischer2021}, instance segmentation~\cite{Fischer2021} and license plate recognition~\cite{Kaiser2021}. 
A common approach to encounter this decrease in performance is to enlarge data sets by data augmentation, i.e., extending them with modified versions of the original images by applying expected distortions synthetically~\cite{Shorten2019}.
\begin{figure}
	\centering
	\begin{tikzpicture}
		\node (imA) at (0,0) {\includegraphics[trim={0 10 0 15},clip,width=.95\linewidth]{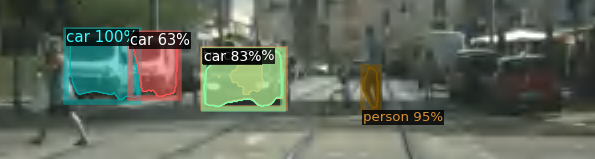}};
		\node [align=center, below=-.1cm of imA.south] (imAcap)  {(a) Baseline};
		\node[align=center, below=-.15cm of imAcap.south] (imC) {\includegraphics[trim={0 10 0 15},clip,width=.95\linewidth]{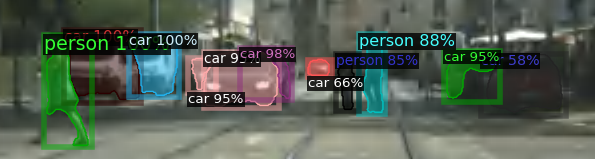}};
		\node [align=center, below=-.1cm of imC.south] (imCcap)  {(b) Our approach};
	\end{tikzpicture}	
	\vspace*{-1.2em}
	\caption{Visual comparison of instance segmentation results on Cityscapes image \texttt{frankfurt\_0000000\_001236} for test data subject to an unknown distortion (here: JPEG2000 at a $\jkQP$ of $\SI{32}{\decibel}$) with and without our proposed approach for unpaired learning of unknown distortions.}
	\vspace*{-1.5em}
	\label{fig:motivation}
\end{figure}

However, not all image and video distortions may be easily modeled and applied to pristine images.
For example, translating images between different camera setups is a more complex problem.
Therefore, first attempts at learning pristine-to-distorted mapping functions have been made~\cite{Chen2020a}. A pristine-to-distorted mapping function translates images from the source domain $\undistDom$ containing pristine, undistorted images to the target domain $\distDom$ containing distorted images. 
Learning mappings between images belonging to different domains is also termed image-to-image translation.

For paired training of image-to-image translation, images from domain \undistDom{} and \distDom{} related by the true pristine-to-distorted mapping function \distFct{} are required. 
In~\cite{Isola2017}, Isola et al. have proposed the \pixpix{} framework for paired image-to-image translation. This method employs adversarial learning, and is shown to be a flexible framework able to learn a large variety of mappings between domains.
Paired image-to-image translation is impossible when regarding \emph{unknown distortions}, since access to the true distortion function \distFct{} is required to obtain the training data.
In~\cite{Chen2020a}, Chen et al. show that their \pixpix{}-based approach is capable of learning distortions from paired data. The visual impression is verified by an evaluation in terms of natural scene statistics. However, the effectiveness of the approach is not investigated with regards to machine vision tasks.
In~\cite{Zhu2017}, Zhu et al. extend \pixpix{} to \cyclegan{} by introducing a cycle-consistency loss. This enables the unpaired learning of mappings between domains in a similar flexible framework.
For unpaired image-to-image translation unrelated images may be taken from domains \undistDom{} and \distDom{} enabling the learning of \emph{unknown distortions}.

In contrast to utilizing image-to-image translation to align data of differing domains on a pixel-level, a domain gap may also be overcome by aligning the training and testing data on feature-level.
However, the advantage of pixel-level approaches is that the alignment of data and the method to solve the machine vision task may be regarded independently.
For this reason, we only regard alignment on pixel-level, i.e., pixel-level domain adaptation, by means of image-to-image translation in this work. 

In this work, we show that with our approach for unpaired learning of unknown distortions, blur, white noise, JPEG coding, JPEG2000 coding, and HEIF coding may be emulated at several levels of distortion.
We compare the performance of instance segmentation in an autonomous driving scenario when adapting the model by means of corresponding true and learned pristine-to-distorted mapping functions. 
In this context, we also perform an extensive benchmark of the impact of image distortions on instance segmentation employing Mask \ac{R-CNN}.

A visual impression of the advantage of employing our proposed unpaired learning of unknown distortions in instance segmentation is given in \autoref{fig:motivation}. In \autoref{fig:motivation}, multiple instances in a distorted image are not recognized by the baseline system trained on pristine data. 
By adapting the instance segmentation to the unknown distortion (here: JPEG 2000 at a $\jkQP$ of $\SI{32}{\decibel}$) with our proposed approach, more instances are recognized.

%% file: texfiles/3_methods.tex
\section{Unsupervised Domain Adaptation for Unknown Distortions}
\label{sec:method}
\acp{DNN} for instance segmentation in autonomous driving are commonly trained on pristine data.
However, in practical applications, the test data is often subject to unknown distortions impeding the performance of the \ac{DNN}.
We propose to overcome this domain shift between the source domain \undistDom{} containing pristine, undistorted data and the target domain \distDom{} subject to an unknown distortion by means of unpaired image-to-image translation.
The pristine-to-distorted mapping function $\distFct : \undistDom \rightarrow \distDom$ represents the unknown mapping from the source to the target domain.
Learning this mapping function from unpaired data enables the emulation of the unknown distortion on the labeled training data. With this unsupervised domain adaptation, the performance in the target domain may be improved without access to labeled training data.

\subsection{Unpaired Learning of Unknown Distortions}
\label{sec:method_ulud}
For unpaired learning of the unknown pristine-to-distorted mapping function \distFct{}, the image-to-image translation technique \cyclegan{} is employed~\cite{Zhu2017}. 
This unpaired approach consists of a generator \generatorForward{} to translate images from the undistorted to the distorted domain $\generatorForward : \undistDom \rightarrow \distDom$ and a generator \generatorBackward{} to translate images from the distorted to the undistorted domain $F : \distDom \rightarrow \undistDom$. 
Employing a cycle-consistency loss, the difference between input images $\myvec{x} \in \undistDom$ and $\myvec{y} \in \distDom$ and their respective translations employing both generators $\hat{\myvec{x}} = \generatorBackward(\generatorForward(\myvec{x}))$ and $\hat{\myvec{y}} = \generatorForward(\generatorBackward(\myvec{y}))$ is minimized.
Furthermore, discriminators \discriminatorUndist{} and \discriminatorDist{} exist to distinguish between true and generated samples of domains \undistDom{} and \distDom{}, respectively.
In our approach, the learned mapping function $\distFctLearned{}$ for the unknown distortion is given by generator \generatorForward{}. The optimal generator $G^*$ is obtained by solving
\begin{equation}
	G^* = \mathrm{argmin}_{G,F,D_X,D_Y} \ \mathcal{L}(G,F,D_X,D_Y).
\end{equation}
For further details on \cyclegan{} and the full objective function $\mathcal{L}(G,F,D_X,D_Y)$, we refer the reader to the original publication and the reference implementation provided by the authors~\cite{Zhu2017}.

With the goal of obtaining one set of parameters to learn different unknown distortions at arbitrary strengths, multiple preliminary experiments were conducted.
We empirically found the setup presented in~\cite{Zhu2017} for the translation between paintings and photos to perform best over a broad range of distortions at different strengths.
These training parameters entail a cycle-consistency loss weighing factor of \num{10}, an identity mapping loss weighing factor of \num{0.5} and a batch size of \num{1}. The models are trained from scratch, employing a constant learning rate of \num{0.0002} during the first \num{100} epochs, and linearly decaying it to zero over the following \num{100} epochs.
The generator networks \generatorForward{} and \generatorBackward{} consist of \num{9} residual blocks. Discriminators \discriminatorUndist{} and \discriminatorDist{} are $70 \times 70$ PatchGANs~\cite{Zhu2017}.
The training data is randomly cropped to a size of $256 \times 256$.

An advantage of our proposed approach is the independence from prior knowledge of the unknown distortions by employing a single set of parameters for a wide range of distortions at arbitrary levels.
The flexibility to adapt to a wide range of distortions at various levels is of particular importance, since commonly images and videos are not only subject to a single distortion, but combinations thereof.

Another benefit of our approach is the unnecessity of additional data sets. We employ the instance segmentation test images from the target domain \distDom{} and the instance segmentation training images from the undistorted source domain \undistDom{} to train the image-to-image translation in an unpaired manner.
The learned mapping \distFctLearned{} is then employed to emulate the unknown distortion on the instance segmentation training data.

\subsection{Adapting Instance Segmentation to Unknown Distortions}
In~\cite{Fischer2021}, \nobreakdash{Fischer} et al. show that robustness against distortion may be obtained either by including the degraded images into the training data, or by fine-tuning a network trained on pristine data on the respective distortion.
Therefore, we choose the same pre-trained Mask \ac{R-CNN}~\cite{He2020} provided in~\cite{Wu2019} and perform a fine-tuning on training data distorted by means of the learned pristine-to-distortion mapping function \distFctLearned{}.

The \detectron{} framework~\cite{Wu2019} is employed to fine-tune Mask \ac{R-CNN}~\cite{He2020} for instance segmentation.
The employed pre-trained model has a ResNet50~\cite{He2016} backbone, and is pre-trained on COCO~\cite{Lin2014} and pristine Cityscapes~\cite{Cordts2016}.
With a batch size of \num{4}, the fine-tuning is performed for a maximum of \num{24000} iterations, to improve the \ac{DNN}'s performance on the target domain. Starting with a learning rate of \num{0.01}, the learning rate is decreased after \num{18000} iterations to \num{0.001}.

%% file: texfiles/4_experiments.tex
%
\section{Experimental Setup}
\label{sec:experiments_setup}
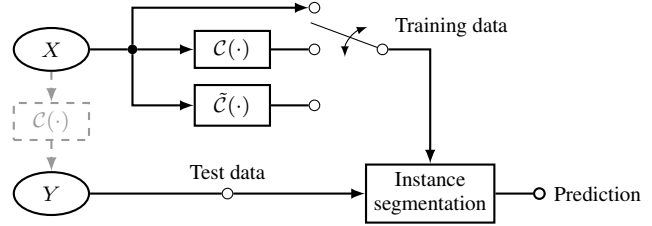
\begin{figure}
	\centering
	\input{images/framework_blockDiag.tex}
	\vspace*{-1.4em}
	\caption{Experimental setup of unsupervised domain adaptation for unknown distortions. Mapping function \distFct{} represents the unknown mapping from undistorted source domain \undistDom{} to distorted target domain \distDom{}. Mapping function \distFctLearned{} is learned by means of unpaired image-to-image translation.}
	\vspace*{-1.2em}
	\label{fig:framework_blockDiag}
\end{figure}
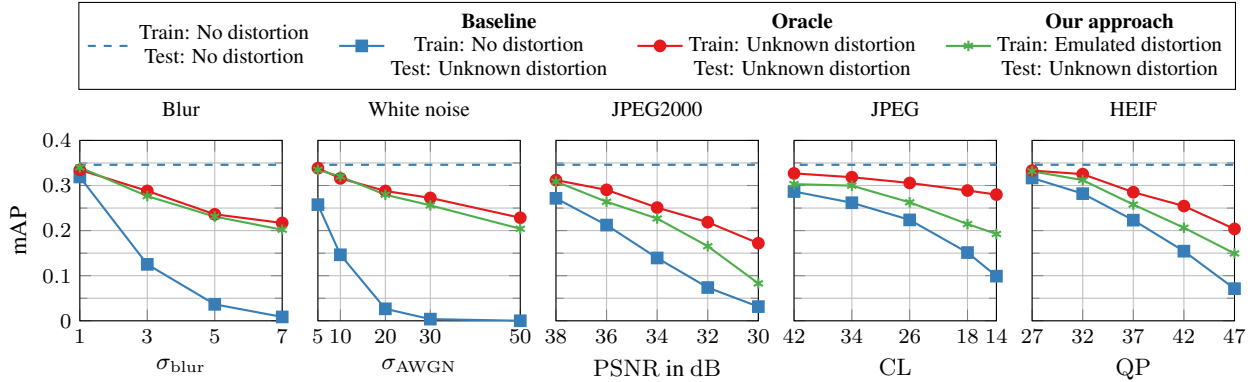
\begin{figure*}[!htb]
	\centering
	\input{images/all_aps_tikzpicture.tex}
	\vspace*{-0.9em}
	\caption{Results of instance segmentation measured as \acf{mAP} over distortion level for various distortion types.}
	\vspace*{-1.1em}
	\label{fig:results_ap}
\end{figure*}
The experimental setup to evaluate our proposed approach is depicted in \autoref{fig:framework_blockDiag}. We employ data from the undistorted source domain \undistDom{} as training data. We regard three different scenarios illustrated by the three different branches: 
\begin{description}[leftmargin=2em]
	\item[Baseline] In the top branch, we directly employ data from the undistorted source domain \undistDom{} for training. Hence, the instance segmentation system is not adapted to the distorted domain \distDom{} and the distortion remains unknown.
	\item[Oracle-based approach] In the middle branch, the true pristine-to-distorted mapping function \distFct{} is employed to adapt the instance segmentation. This represents the best-case scenario. However, since our proposed approach is meant to adapt instance segmentation to unknown distortions of arbitrary strength, this \emph{oracle-based} approach is unattainable in practice.
	\item[Our proposed approach] As detailed in \autoref{sec:method_ulud}, we learn a mapping function \distFctLearned{} from pristine images to an unknown distortion by means of unpaired image-to-image translation. We emulate the unknown distortion on the training data employing \distFctLearned{} to adapt instance segmentation to the unknown distortion.
\end{description}

The evaluation is carried out employing test data from the distorted domain \distDom{}. This distorted test data is obtained by applying the true distortion mapping function \distFct{} to the Cityscapes validation split.
The performance of Mask \ac{R-CNN} is evaluated in terms of \ac{mAP} and reported over the level of distortion. The \ac{mAP} is calculated as described in~\cite{Cordts2016}, by first calculating the \ac{AP} for each class for various overlaps, and then taking the mean over all classes. 

We employ the Cityscapes dataset introduced in~\cite{Cordts2016}. We choose this data set, since it contains relatively little distortions in comparison to other data sets for autonomous driving. 
For each distortion employed in this work, 
we degrade all splits of the Cityscapes data set in order to train the unpaired image-to-image translation as well as the instance segmentation.

To show the flexibility of our proposed approach, we choose a diverse set of distortions at different levels of distortion, namely blur, \acf{AWGN} and different image compression techniques. We show the applicability of our approach for compression with a fixed block size (JPEG), with adaptive block-size (HEIF), and wavelet-based coding (JPEG2000).

The distortion level of \emph{blur} is denoted by standard deviation \blurQP{} of the two-dimensional symmetric Gaussian kernel, with which the pristine image is convolved.
For \emph{white noise}, random values are drawn independently for each pixel from a zero-mean Gaussian distribution with standard deviation \noiseQP{} and added to the undistorted image.
For \emph{JPEG} and \emph{JPEG2000} en- and decoding methods provided by \imageMagick{}~\cite{imagemagick} are employed. The JPEG quality is varied by means of \acf{CL}, with zero leading to the strongest and \num{100} to the lowest level of distortion.
The quality of JPEG2000 is controlled by the \acf{PSNR} in $\si{\decibel}$ between pristine and coded image.
\emph{HEIF} en- and decoding is performed employing \libheif{}~\cite{libheif}. The distortion level is controlled by the \acf{QP} ranging from zero to \num{51}. With a \ac{QP} of \num{51} the strongest distortions are introduced.

\section{Results and Discussion}
\label{sec:experiments}
\sisetup{
	round-mode = places,
	round-precision = 2,
}
\begin{table}
	\centering
	\input{csvs/ap_averages/ap_averages.tex}
	\vspace*{-0.5em}
	\caption{Average \acf{mAP} gains over baseline model without adaption to distortions, when employing an oracle and our proposed approach. The last column shows the difference between our proposed approach and the oracle-based approach.}
	\vspace*{-1.0em}
	\label{tab:avgAP}
\end{table}
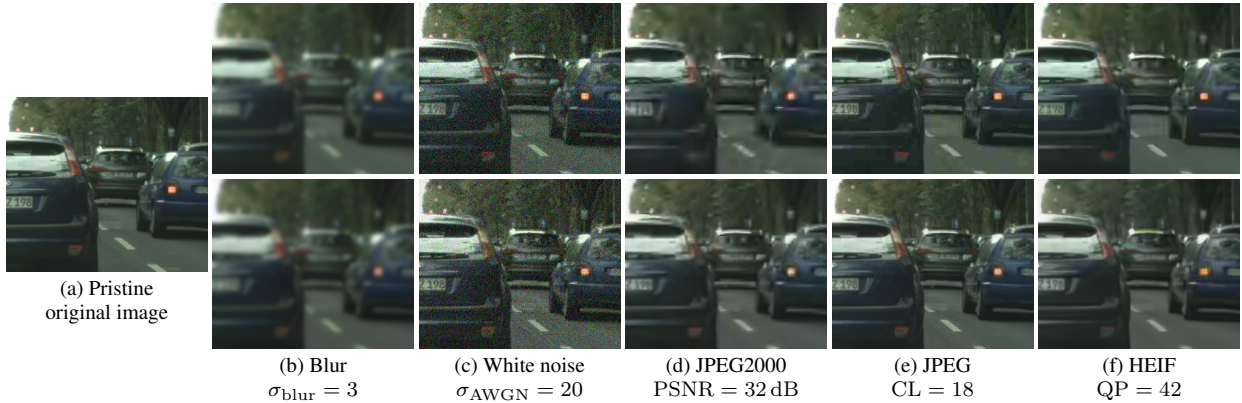
\begin{figure*}[htb]
	\centering
	\input{images/vis_examples.tex}
	\vspace*{-1.0em}
	\caption{Visual examples of applying true (top row) and learned (bottom row) distortion mapping function to Cityscapes image \texttt{bremen\_000117\_000019} for various distortion types. \textit{(Best to be viewed enlarged on a monitor.)}}
	\vspace*{-0.9em}
	\label{fig:vis_examples}
\end{figure*}
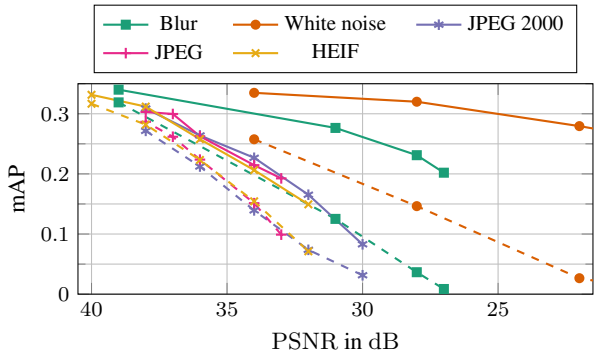
\begin{figure}[htb]
	\centering
	\input{images/map_over_psnr.tex}
	\vspace*{-1.0em}
	\caption{Results of instance segmentation measured as \acf{mAP} over \acf{PSNR} for various distortion types. Dashed plots denote results obtained with the baseline, solid lines denote results obtained with our proposed approach.}
	\vspace*{-1.5em}
	\label{fig:map_over_psnr}
\end{figure}
The experimental results in terms of \ac{mAP} over the level of distortion are shown in \autoref{fig:results_ap}.
The distortion levels are sorted so that the leftmost value corresponds to the lowest and the rightmost value to the highest distortion.
The blue dashed line shows the \ac{mAP} obtained when testing the pre-trained baseline model on pristine data. 
With this setup the best results are obtained reaching an \ac{mAP} of \num{0.3458739657843787}.
For all solid plots, the test data is distorted with the respective true distortion function \distFct{}.
The blue solid line depicts the results for the baseline scenario without fine-tuning of the model.
For all types of distortion the \ac{mAP} decreases with the increasing level of distortion.
The red solid line depicts the results of the oracle with knowledge of the true distortion function \distFct{}.
Here each distortion at each strength is regarded as its own target domain, i.e., for each red point a specific instance segmentation model was adapted.
For very low distortion like a blur of $\blurQP=1$, the difference between the pre-trained baseline approach and the adapted oracle-based approach is very small. For stronger degradations, the mAP is increased by the specialization on the regarded distortion.

The green solid line depicts the results for our proposed unpaired learning of unknown distortions.
For all distortions at all distortion levels, the mAP is increased in comparison to the baseline, i.e., the case without adaptation to the unknown distortion.
For blur and white noise, mAPs very close to the oracle-based approach are reached. 
For the different image coding techniques, the obtained mAP is also quite close to the mAP achieved by the oracle.  
For stronger distortions, the gap between mAPs obtained by the true and the learned distortion function becomes larger.

In \autoref{tab:avgAP}, the gains in \ac{mAP} over the baseline are averaged over the level of distortion.
It can be seen that we achieve results comparable to the oracle-based approach.
The largest average gain in \ac{mAP} of \num{0.19202720733033435} is observed for white noise.
The average gain in mAP for white noise and blur achieved with our proposed approach is only \num{0.01} lower than the oracle-based approach.
For image compression with our proposed approach average gains ranging from \num{0.04} to \num{0.06} are achieved. The gains obtained with the oracle are between \num{0.03} to \num{0.05} higher.

Visual examples for one distortion level of each distortion type are shown in \autoref{fig:vis_examples}. 
In \autoref{fig:vis_examples}(a), a pristine example image is depicted. 
In \autoref{fig:vis_examples}(b)-(f), this image is distorted by the true pristine-to-distorted mapping \distFct{} in the top row, and the the learned mapping \distFctLearned{} in the bottom row.
For blur and white noise it can be seen that the learned distortion is visually very similar to the true distortion.
In the case of the different image compression techniques, the learned and true distortions are also visually close. However, not all types of occurring artifacts can be reproduced yet.

For a better comparison in terms of objective image quality, the results of the instance segmentation measured as \ac{mAP}, are shown as a function of \ac{PSNR} in \autoref{fig:map_over_psnr}. 
Therefore, the mean \ac{PSNR} was computed for each distortion level of each distortion type. The dashed plots denote the results obtained with the baseline, the solid plots represent the results of our proposed approach.
In the baseline approach, for the same distortion level in terms of \ac{PSNR}, the highest mAPs are observed for white noise, followed by blur. 
All regarded image compression techniques have a similar influence, resulting in a lower mAP. 
Our proposed unpaired learning of unknown distortions improves the instance segmentation results for all \ac{PSNR} levels. The sensitivity towards the different distortion types remains the same.

It can be seen, that the employed instance segmentation model exhibits a strong sensitivity towards various image distortion. 
The instance segmentation seems to be very sensitive towards image compression, while exhibiting more resiliency toward blur and white noise.
As expected, for the \emph{unknown distortion}, i.e., model trained on undistorted source domain \undistDom{} and tested on distorted target domain \distDom{}, the lowest mAPs are achieved.
With the regarded \emph{oracle-based approach}, i.e., model specialized and tested on distorted target domain \distDom{}, the highest mAPs are obtained.
With our proposed \emph{unpaired learning of unkown distortions} results comparable to the \emph{oracle-based approach} are obtained. 
Moreover, we can achieve mAPs for blur and white noise closely matching the oracle-based approach. 
For the different image coding techniques, especially at a higher level of distortion there is still room for improvement.
A potential reason for the larger gap in image coding techniques is the inability of the employed generator network to reproduce block artifacts.

%% file: images/framework_blockDiag.tex
\begin{tikzpicture}
	\node[source, minimum width=1cm, black, thick] (undistDom) {\undistDom{}};
	\node[system, below=.4cm of undistDom, minimum width=1cm, dashed, mygray, thick] (distFct) {$\distFct(\cdot)$};
	\node[source, minimum width=1cm, black, below=.4cm of distFct, thick] (distDom) {\distDom{}};
	
	\draw[->, dashed, mygray, thick] (undistDom) -- (distFct);
	\draw[->, dashed, mygray, thick] (distFct) -- (distDom);
	
	\node[syssplit, right=.5cm of undistDom, thick] (split) {};
	\draw[-, thick] (undistDom) -- (split);
	
	\node[system,right=.75cm of split,align=center, minimum width=1cm, thick](oracle){$\distFct(\cdot)$};
	\draw[->, thick] (split) -- (oracle);
	\node[syscon,right=.5cm of oracle](oracleCon){};
	\draw[-, thick] (oracle) -- (oracleCon);
	
	\node[system,below=.2cm of oracle,align=center, minimum width=1cm, thick](distFctLearned){$\distFctLearned(\cdot)$};
	\draw[->, thick] (split) |- (distFctLearned);
	\node[syscon,right=.5cm of distFctLearned](distFctLearnedCon){};
	\draw[-, thick] (distFctLearned) -- (distFctLearnedCon);
	
	\node[syscon,above=.4cm of oracleCon](noFTCon){};
	\draw[->, thick] (split) |- (noFTCon);
	
	\node[syscon,right=.75cm of oracleCon, label={[align=left]above right:{Training data}}](casesEnd){};

	\node[syscon,right=1.75cm of distDom, label={[align=center]above:{Test data}}](testData){};
	\node[system,right=1.75cm of testData,align=center, minimum width=1cm, thick](instSeg){Instance\\segmentation};
	\draw[->, thick] (casesEnd) -| (instSeg) {};
	\draw[-, thick] (distDom) -- (testData) {};
	\draw[->, thick] (testData) -- (instSeg) {};
	
	\node[syscon, right=.5cm of instSeg, label={[align=center] right:{Prediction}}, thick] (output) {};
	\draw [thick] (instSeg) -- (output);
	
	\coordinate[above left= .3 and .9 of casesEnd] (switch);
	\draw (casesEnd) -- (switch);
	\draw[<->] (casesEnd) ++(-0.5, -0.1) to[out=90, in=-160] ++(.3, .4);

\end{tikzpicture}

%% file: images/all_aps_tikzpicture.tex
\definecolor{col1}{RGB}{55,126,184}
\definecolor{col2}{RGB}{228,26,28}
\definecolor{col3}{RGB}{77,175,74}
\definecolor{col4}{RGB}{152,78,163}
\definecolor{col5}{RGB}{255,127,0}
\begin{tikzpicture}
 	\begin{groupplot}[
 		group style={
 			group size=5 by 1,
 			group name=allAps,
 			x descriptions at=edge bottom,
 			y descriptions at=edge left,
 			horizontal sep=1.5em
			},
 		width=.24\textwidth,
		ymin=0,
		ymax=0.40,
		ytick={0,0.1,...,0.4},
		ylabel={mAP},
		y=6cm,
		xtick=data,
		enlarge x limits=false,
		grid=both,
		minor x tick num=0,
		minor y tick num=1,
 		/tikz/font=\footnotesize
 		]
 		\input{images/blur_gauss_ap_group.tex}
		\input{images/noise_gauss_ap_group.tex}
		\input{images/j2k_ap_group.tex}
		\input{images/jpeg_ap_group.tex}
		\input{images/heif_ap_group.tex}
	\end{groupplot}
\end{tikzpicture}

%% file: csvs/ap_averages/ap_averages.tex
\begin{tabular}{cccc}
\toprule
  Distortion &                     Oracle &                 Our approach &                   Difference \\
\midrule
        Blur &   \num{0.1466361305005603} &   \num{0.1401212389039071} &  \num{-0.006514891596653194} \\
 White noise &  \num{0.20175787473191625} &  \num{0.19202720733033435} &  \num{-0.009730667401581905} \\
    JPEG2000 &  \num{0.10312259177096875} &  \num{0.06387885710574537} &   \num{-0.03924373466522338} \\
        JPEG &  \num{0.09942386653142507} &  \num{0.04999016849912449} &   \num{-0.04943369803230058} \\
        HEIF &  \num{0.07083362302796473} &  \num{0.04188097731349663} &  \num{-0.028952645714468095} \\
\bottomrule
\end{tabular}

%% file: images/vis_examples.tex
\begin{tikzpicture}
	\node (imA) at (0,0) {\includegraphics[trim={40 70 35 105},clip,width=.15\linewidth]{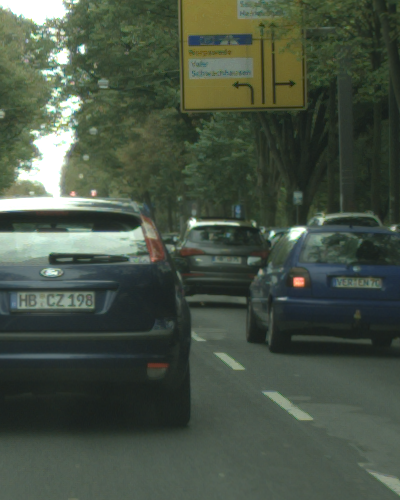}};
	\node[align=center, right=-.15cm of imA.north east] (imB) {\includegraphics[trim={40 75 35 105},clip,width=.15\linewidth]{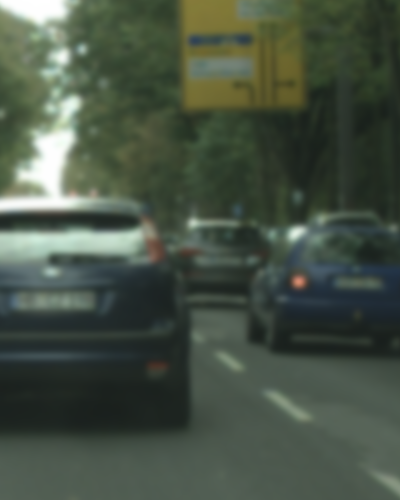}};
	\node[align=center, right=-.15cm of imB.east] (imC) {\includegraphics[trim={40 75 35 105},clip,width=.15\linewidth]{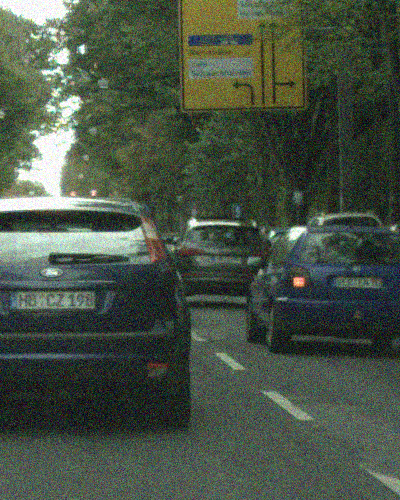}};
	\node[align=center, right=-.15cm of imC.east] (imD) {\includegraphics[trim={40 75 35 105},clip,width=.15\linewidth]{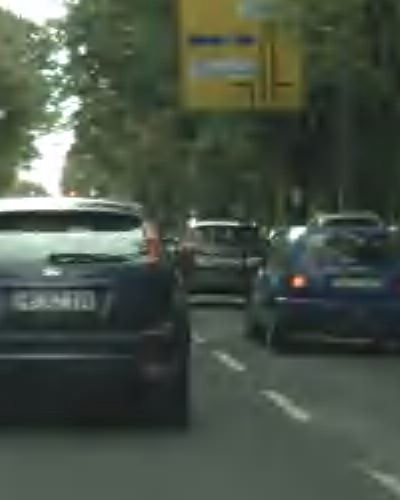}};
	\node[align=center, right=-.15cm of imD.east] (imE) {\includegraphics[trim={40 75 35 105},clip,width=.15\linewidth]{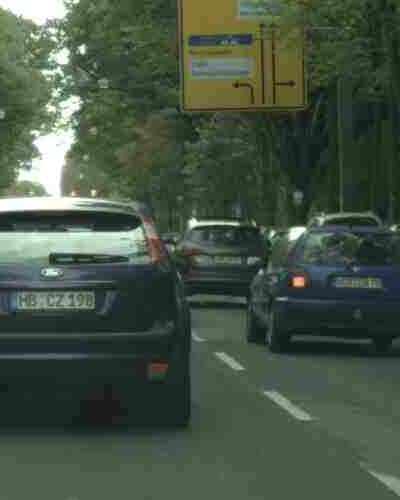}};
	\node[align=center, right=-.15cm of imE.east] (imF) {\includegraphics[trim={40 75 35 105},clip,width=.15\linewidth]{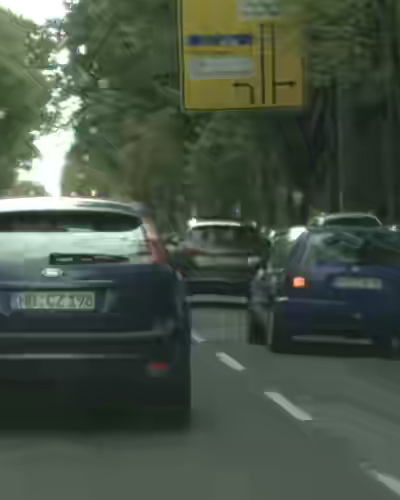}};
	
	\node[align=center, below=-.15cm of imB.south] (imB) {\includegraphics[trim={40 75 35 105},clip,width=.15\linewidth]{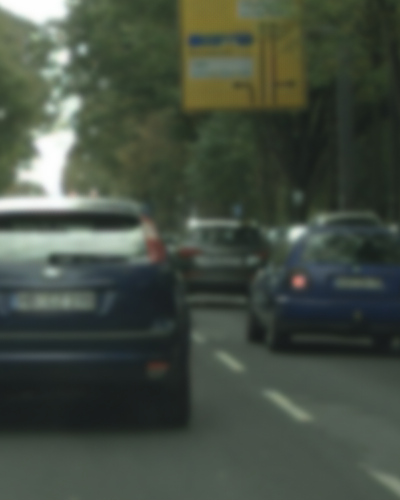}};
	\node[align=center, below=-.15cm of imC.south] (imC) {\includegraphics[trim={40 75 35 105},clip,width=.15\linewidth]{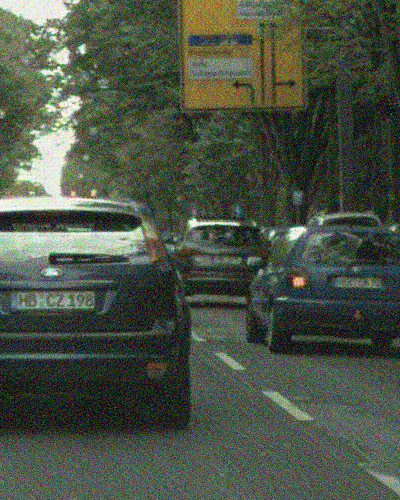}};
	\node[align=center, below=-.15cm of imD.south] (imD) {\includegraphics[trim={40 75 35 105},clip,width=.15\linewidth]{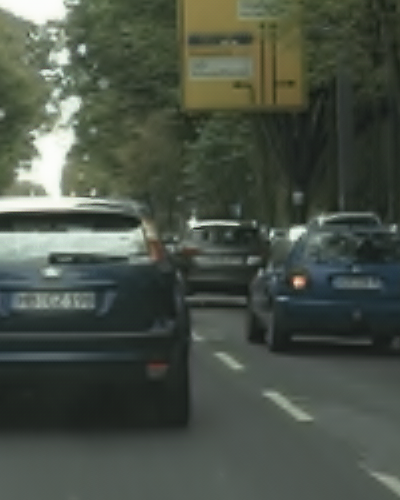}};
	\node[align=center, below=-.15cm of imE.south] (imE) {\includegraphics[trim={40 75 35 105},clip,width=.15\linewidth]{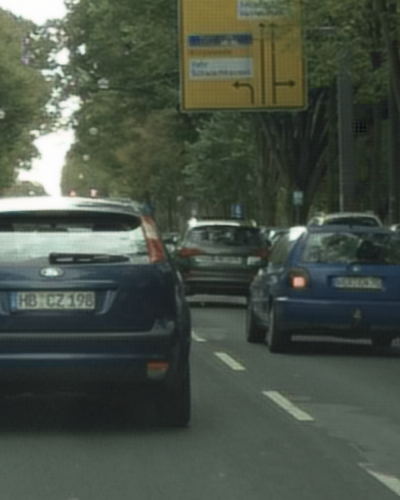}};
	\node[align=center, below=-.15cm of imF.south] (imF) {\includegraphics[trim={40 75 35 105},clip,width=.15\linewidth]{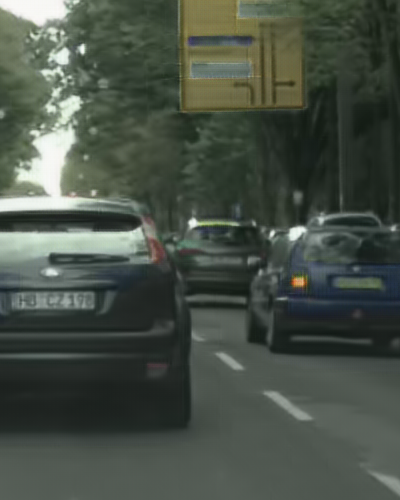}};
	
	\node [align=center, below=-.1cm of imA.south] (imAcap)  {(a) Pristine\\original image};
	\node [align=center, below=-.1cm of imB.south] (imBcap)  {(b) Blur\\$\blurQP = 3$};
	\node [align=center, below=-.1cm of imC.south] (imCcap)  {(c) White noise\\$\noiseQP = 20$};
	\node [align=center, below=-.1cm of imD.south] (imDcap)  {(d) JPEG2000\\$\jkQP = \SI[round-mode=places,round-precision=0]{32}{\decibel}$};
	\node [align=center, below=-.1cm of imE.south] (imEcap)  {(e) JPEG\\$\jpegQP = 18$};
	\node [align=center, below=-.1cm of imF.south] (imFcap)  {(f) HEIF\\$\heifQP = 42$};
\end{tikzpicture}

%% file: images/map_over_psnr.tex
\definecolor{col1}{RGB}{27,158,119}
\definecolor{col2}{RGB}{217,95,2}
\definecolor{col3}{RGB}{117,112,179}
\definecolor{col4}{RGB}{231,41,138}
\definecolor{col5}{RGB}{230,171,21}
\begin{tikzpicture}
\begin{groupplot}[
	group style={
		group size=1 by 1,
		group name=allApsPSNR,
		x descriptions at=edge bottom,
		y descriptions at=edge left,
		horizontal sep=1.5em
	},
	width=.98\linewidth,
	ymin=0,
	ymax=0.35,
	ytick={0,0.1,...,0.4},
	ylabel={mAP},
	y=8cm,
	enlarge x limits=false,
	grid=both,
	minor x tick num=0,
	minor y tick num=1,
	/tikz/font=\footnotesize
	]
\nextgroupplot[
xlabel={\jkQP{} in $\si{\decibel}$},
xmin=21.5,
xmax=40.5,
x dir=reverse,
legend style={
	at={(0.5,1.05)},
	anchor=south,
	align=center,
	legend columns=3,
	/tikz/every even column/.append style={column sep=1em},
	fill=none
},
]
%
\addplot[dashed, thick, col1, mark=square*, mark size=1.5, mark options={solid},forget plot]table[x=test_qp, y=ap, col sep=comma]{csvs/ap_used_onlyUsed_qps_as_psnr/blur_gauss_realDegrad_noFT_testDistorted_ap.csv};
\addplot[thick, col1, mark=square*, mark size=1.5, mark options={solid}]table[x=test_qp, y=ap, col sep=comma]{csvs/ap_used_onlyUsed_qps_as_psnr/blur_gauss_cycleganDegrad_FT_testDistorted_ap.csv};
\addlegendentry{Blur}
%
\addplot[dashed, thick, col2, mark=*, mark size=1.5, mark options={solid},forget plot]table[x=test_qp, y=ap, col sep=comma]{csvs/ap_used_onlyUsed_qps_as_psnr/noise_gauss_realDegrad_noFT_testDistorted_ap.csv};
\addplot[thick, col2, mark=*, mark size=1.5, mark options={solid}]table[x=test_qp, y=ap, col sep=comma]{csvs/ap_used_onlyUsed_qps_as_psnr/noise_gauss_cycleganDegrad_FT_testDistorted_ap.csv};
\addlegendentry{White noise}
%
\addplot[dashed, thick, col3, mark=asterisk, mark size=2, mark options={solid},forget plot]table[x=test_qp, y=ap, col sep=comma]{csvs/ap_used_onlyUsed_qps_as_psnr/j2k_realDegrad_noFT_testDistorted_ap.csv};
\addplot[thick, col3, mark=asterisk, mark size=2, mark options={solid}]table[x=test_qp, y=ap, col sep=comma]{csvs/ap_used_onlyUsed_qps_as_psnr/j2k_cycleganDegrad_FT_testDistorted_ap.csv};
\addlegendentry{JPEG 2000}
%
\addplot[dashed, thick, col4, mark=+, mark size=2, mark options={solid},forget plot]table[x=test_qp, y=ap, col sep=comma]{csvs/ap_used_onlyUsed_qps_as_psnr/jpeg_realDegrad_noFT_testDistorted_ap.csv};
\addplot[thick, col4, mark=+, mark size=2, mark options={solid}]table[x=test_qp, y=ap, col sep=comma]{csvs/ap_used_onlyUsed_qps_as_psnr/jpeg_cycleganDegrad_FT_testDistorted_ap.csv};
\addlegendentry{JPEG}
%
\addplot[dashed, thick, col5, mark=x, mark size=2, mark options={solid},forget plot]table[x=test_qp, y=ap, col sep=comma]{csvs/ap_used_onlyUsed_qps_as_psnr/heif_realDegrad_noFT_testDistorted_ap.csv};
\addplot[thick, col5, mark=x, mark size=2, mark options={solid}]table[x=test_qp, y=ap, col sep=comma]{csvs/ap_used_onlyUsed_qps_as_psnr/heif_cycleganDegrad_FT_testDistorted_ap.csv};
\addlegendentry{HEIF}
\end{groupplot}
\end{tikzpicture}

%% file: texfiles/5_conclusion.tex
\section{Conclusion}
\label{sec:conclusion}
In this work we propose to learn the mapping from pristine data to data subject to an unknown distortion. 
We show that this mapping may be learned with a fixed setup for a wide range of distortions and for a diverse set of distortion levels.
This learned mapping is then employed to emulate unknown distortions on pristine, labeled training data for instance segmentation.
By fine-tuning instance segmentation with this training data, we achieve results that are comparable to an oracle with knowledge of the true pristine-to-distorted mapping.
The largest average gain in \ac{mAP} of \num{0.19202720733033435} is obtained for white noise. 
For image coding and blur we achieve average gains in \ac{mAP} between \num{0.04} and \num{0.14}.
Our approach has the advantage that neither prior knowledge of the distortion nor additional data sets are required.
Furthermore, we performed an extensive benchmark on the influence of image distortions on instance segmentation.

In future work, the difference between specialization of \acp{DNN} to certain degradations of certain strengths could be compared to a model trained on various degradations.
Furthermore, an evaluation in terms of visual closeness of the emulated and true distortions may be conducted.
Applying our proposed approach to more complex combinations of distortions could be performed to investigate the generalization capabilities of our approach.